# Triple Motion Estimation and Frame Interpolation based on Adaptive Threshold for Frame Rate Up-Conversion

Hanieh Naderi and Mohammad Rahmati, *Member, IEEE*

*Abstract*—In this paper, we propose a novel motion-compensated frame rate up-conversion (MC-FRUC) algorithm. The proposed algorithm creates interpolated frames by first estimating motion vectors using unilateral (jointing forward and backward) and bilateral motion estimation. Then motion vectors are combined based on adaptive threshold, in order to creates high-quality interpolated frames and reduce block artifacts. Since motion-compensated frame interpolation along unilateral motion trajectories yields holes, a new algorithm is introduced to resolve this problem. The experimental results show that the quality of the interpolated frames using the proposed algorithm is much higher than the existing algorithms.

*Index Terms*—Frame rate up-conversion, frame interpolation, motion estimation, motion compensation.

## I. INTRODUCTION

IN computer video displays, television, and movie data, the frame rate is the number of frames that are displayed per second. The higher the number of frames displaying per second, the smoother the video playback appears to the user [1]. In order to improve the temporal quality and jerkiness artifacts in video and the hold-type display, such as liquid crystal display (LCD), frame rate up-conversion (FRUC) interpolates new frames between original frames and increases the number of frames [2]. FRUC is also used in video compression for reconstructing the skipped frames [3].

FRUC algorithms are generally classified into two methods. The first method generate the interpolated frames using frame repetition and frame averaging from the pixels in the previous and the next frame. The methods in this categories are used because of their simplicity and low complexity. However, these simple methods produce insufficient results that lead motion blurring and ghost artifacts in the interpolated frame [4], [5]. The second method is introduced to solve this problem and is called motion-compensated frame rate up-conversion (MC-FRUC) [6].

MC-FRUC consists of two steps: motion estimation (ME) and motion-compensated interpolation (MCI). Motion estimation tracks motion trajectories between the previous and the next frame and estimates motion vectors. Motion-compensated interpolation uses these motion vectors to produce interpolated frames. The quality of interpolate frames clearly depend on motion estimation and motion-compensated interpolation.

Several motion estimation algorithms such as block matching algorithm (BMA) [7], three step search [8], 2D logarithmic search [9] and diamond search [10] are proposed. Among the noted algorithms, block matching algorithm is the most popular algorithms. Although it is more complex than other fast algorithms, but it can find the most accurate motion in compare to others. Moreover, architecture of [11] can implement the high-speed processing of block matching algorithm by parallel computation.

The traditional block matching algorithm (also known as unilateral motion estimation [12]-[19]), divides one of two neighboring frames into non-overlapped blocks and estimates motion vectors of each block by calculating the sum of absolute difference (SAD) between non-overlapped block in the first frame and each possible block in the search range in the other frame and finally generates interpolated frame along the motion vectors. Unilateral method can creates holes and overlaps in some pixels in the interpolated frame, due to no motion vector and multiple motion vector, respectively. Hence, different algorithms are proposed to handle these problems. For example, to handle overlaps, simple algorithms such as averaging and overwriting on the overlapped pixels are used in [6], [12]. To fill holes, a median filter is used in [18]. In [6], [12] holes are covered by the pixel values from the previous or the next frame.

To solve the problems of overlaps and holes, bilateral motion estimation method is proposed. Bilateral motion estimation [20]-[25] divides the frame to be interpolated into non-overlapped blocks before it is really created and then estimates a motion vector using the temporal symmetry between blocks of the previous and the next frames in the search range around initial block position, which pass through block in the interpolated frame. Since bilateral motion estimation assign one motion vector to each non-overlapped block, overlaps and holes are not generated. However, there is essential problem in existing bilateral methods, when the sum of absolute difference between each two blocks of previous and next frames within the search range is obtained. Motion trajectory of two blocks which have the minimum SAD is taken as the true motion vector. Some of motion vectors may be inaccurate in the uniform or periodic regions. So, if the obtained motion vectors are directly employed for motion compensated interpolation, visual artifacts are observed in the interpolated frames. Hence, to obtain more accurate motion vectors, motion vector smoothing algorithms are proposed [26]-[28], to correct the false motion vectors to enhance the interpolated frame quality.

In this paper, we propose a novel MC-FRUC algorithm that employs unilateral and bilateral methods simultaneously, to benefit the advantages of both methods. The proposed algorithm, first performs bilateral motion estimation and then using vector median filter to obtain more accurate motion vectors. Then overlapped block motion compensation (OBMC) [29], [30] is employed to reduce block artifacts. Second, with combination of forward and backward motion estimation, reduces the number of holes in unilateral motion estimation. To remove the residual holes, the proposed algorithm uses the interpolated frame which obtained from bilateral method. Finally, two interpolated frames which obtained from bilateral and unilateral methods are combined based on adaptive threshold to obtain a higher quality interpolated frame. Fig. 1 shows the block diagram of the proposed MC-FRUC algorithm.

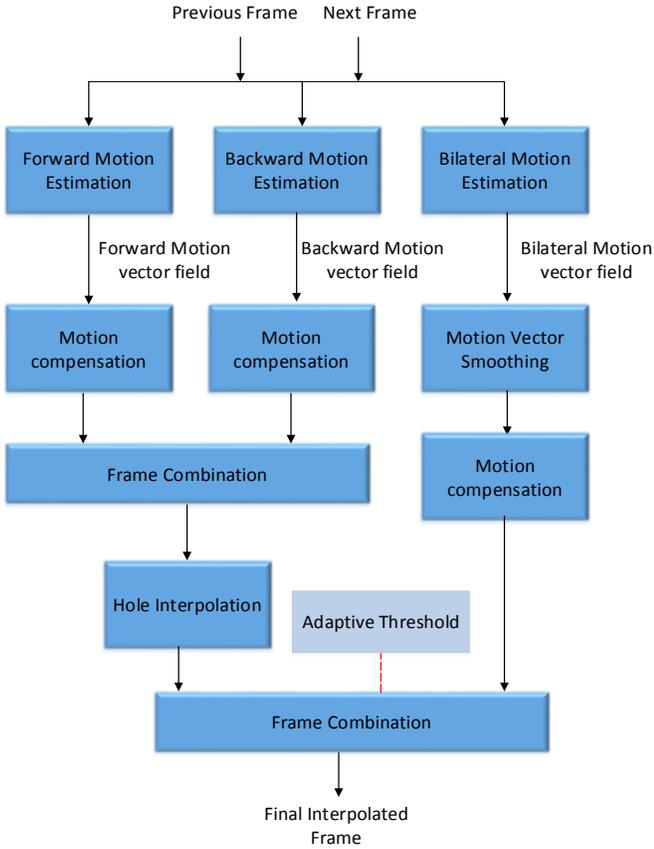

Fig. 1. Block diagram of the proposed MC-FRUC algorithm.

The rest of this paper is organized as follows. Section II describes the proposed algorithm and Section III, considers experimental results. Finally, Section IV concludes this paper.

## II. PROPOSED MC-FRUC ALGORITHM

The proposed algorithm uses the unilateral and bilateral methods simultaneously, to benefit advantages of both methods. Using bilateral method, we solve holes and overlaps problem, but temporal symmetry problem remains. Then, vector median filter (VMF) [31] and overlapped block motion compensation (OBMC) [30] is used to solve the recent problem.

When the unilateral method fails to yield acceptable results, due to holes and overlap areas, the proposed algorithm uses a combinations of forward and backward motion estimation to reduce the number of holes. Finally, a bilateral method is used to remove the remaining holes.

In the proposed algorithm, two interpolated frames which obtained from bilateral and unilateral methods are combined based on adaptive threshold to improve the quality of interpolate frame. The details are given in the following subsections.

### A. Motion Estimation

The bilateral and unilateral motion estimation methods are simultaneously used to estimate the initial motion vector fields.

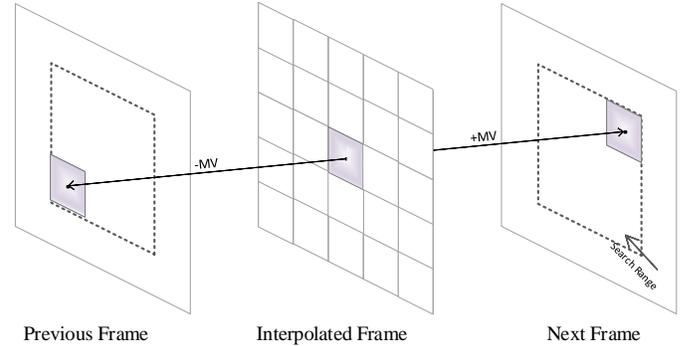

Fig. 2. Bilateral motion estimation.

In bilateral motion estimation, firstly we divide the frame to be interpolated into non-overlapped blocks before it is really created. Then we select a linear trajectory between the previous and the next frames passing at the center of the each block in the interpolated frame as shown in Fig. 2. We use sum of absolute difference (SAD) as the motion vector validity measure that is defined as,

$$SAD = \sum_{i=1}^{m}\sum_{j=1}^{n}\left|f_p(i,j) - f_n(i,j)\right| \qquad (1)$$

where $m$ and $n$ denote the horizontal length and vertical length of candidate blocks, respectively. The $f_p$ denotes the pixel value of the candidate block in the previous frame and $f_n$ denotes the pixel value of the candidate block in the next frame and *(i,j)* represents the pixel location in the candidate block. The search range is limited to a small displacement around the initial block position. Motion vectors are found in accordance with the smallest SAD of two candidate blocks of previous and next frames in the search range. The motion vectors for each block in the interpolated frame is obtained. These motion vectors between the interpolated frame and previous and next frames are symmetric. Bilateral motion vector fields can be obtained by this motion estimation method.

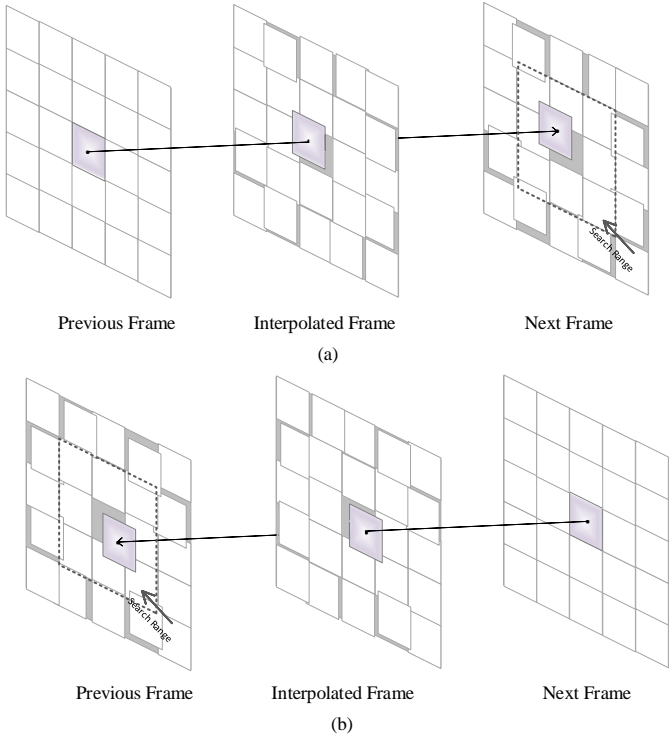

Fig. 3. Unilateral motion estimation (forward motion estimation and backward motion estimation.) (a) Forward motion estimation. (b) Backward motion estimation.

In unilateral motion estimation we use forward motion estimation and backward motion estimation. In forward motion estimation, firstly we divide previous frame into non-overlapped blocks and then define a search range around the initial block. Each block in the previous frame is compared with all of the blocks in the search range in the next frame. The block which have the least SAD with block in the previous frame is chosen and its motion vector is selected. The backward motion estimation derived in the same way. As shown in Fig. 3, overlaps and holes are created according motion vector trajectory. The gray regions in the interpolated frame show the holes and the overlapped area between the small squares in the interpolated frame are called overlaps.

Forward and backward motion vector fields can be obtained by forward and backward motion estimation methods.

*B. Motion Vector Smoothing*

As noted in the previous section, bilateral motion vector fields are obtained by bilateral motion estimation. The motion vector field cannot exactly indicate motion of objects, since the property of the temporal symmetry of bilateral motion estimation leads to estimate of false motion vectors. The false motion vectors cause extremely degraded frame quality. We smooth motion vector field and remove outliers in the motion vector field, using vector median filtering. In fact, each candidate motion vector is replaced by a median of candidate motion vector and its eight neighboring motion vectors.

*C. Motion Compensated Interpolation*

Once the final motion vector field is obtained, the interpolated frame can be obtained using bilateral motion-compensated interpolation algorithm as follows,

$$f_{bi}(i,j) = \frac{1}{2}(f_p(i+mv_x, j+mv_y) + f_n(i-mv_x, j-mv_y)), \quad (2)$$

where $f_{bi}$ denotes the pixel value of the bilateral interpolated frame, and $mv = (mv_x, mv_y)$ denotes motion vector after motion vector smoothing.

After constructing the interpolated frame, overlapped block motion compensation (OBMC) is finally applied to reduce block artifacts in the interpolated frame. As shown in Fig. 4, each block in the interpolated frame is considered an enlarged block which overlap with neighboring blocks.

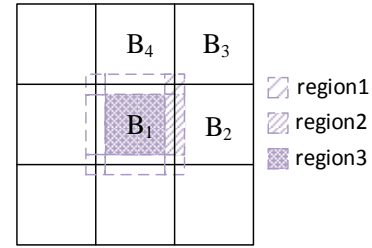

Fig. 4. Overlapped block motion compensation.

Consider region 1 in Fig. 4 which overlaps four blocks *($B_1$, $B_2$, $B_3$, $B_4$)*. To obtain a pixel $f_b(i_1, j_1)$ in region 1, the following is used,

$$f_{bi}(i_1, j_1) = \frac{1}{8}(\sum_{i=1}^{4}(f_p(i_1+mvx_i, j_1+mvy_i) + f_n(i_1-mvx_i, j_1-mvy_i)) \quad (3)$$

where $mv_i = (mvx_i, mvy_i)$ denotes for motion vectors of four neighboring blocks *($B_1$, $B_2$, $B_3$, $B_4$)* in region 1.

Any pixel in region 2, which overlaps two blocks *(B1, B2)*, is obtained by,

$$f_{bi}(i_1, j_1) = \frac{1}{4}(\sum_{i=1}^{2}(f_p(i_1+mvx_i, j_1+mvy_i) + f_n(i_1-mvx_i, j_1-mvy_i)) \quad (4)$$

where $mv_i = (mvx_i, mvy_i)$ denotes the motion vectors of two neighboring blocks $(B_1, B_2)$ in region 2. Similarly, pixel values in region 3 can be obtained.

In unilateral method, forward interpolated frame, $f_f$, is created using the forward motion vector field obtained in section A. The forward interpolated frame, $f_f$, can be obtained using unilateral motion-compensated interpolation algorithm as follows,

$$f_f(i + \frac{mv_{xf}}{2}, j + \frac{mv_{yf}}{2}) = \frac{1}{2}(f_p(i,j) + f_n(i+mv_{xf}, j+mv_{yf})) \quad (5)$$

where $mv = (mv_{xf}, mv_{yf})$ denotes motion vector which is associated with forward motion vector field. The backward interpolated frame would be derived in the same way.

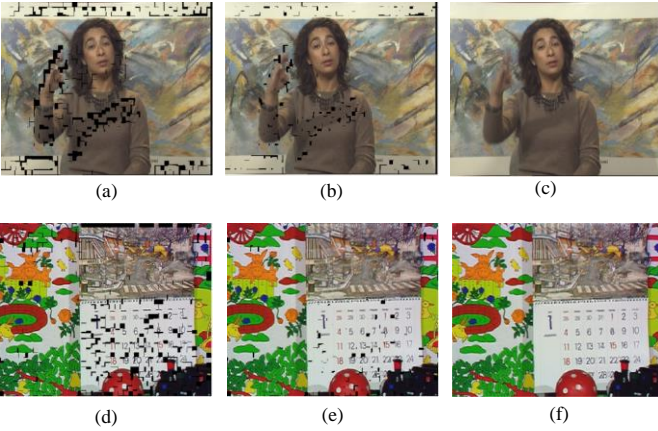

(a) (b) (c)
(d) (e) (f)

Fig. 5. Comparison of interpolated frames from sequences *Silent* and *Mobile* before (a) and (d) and after handle holes (b), (e), (c) and (f). Frames (a) and (d) are forward interpolated frames only. Frames (b) and (e) are combination of forward and backward interpolated frames. Frames (c) and (f) are interpolated frames which obtained after filling holes with proposed algorithm.

In forward and backward interpolated frames, the unknown pixels (holes) cannot be interpolated using motion compensation because the motion for these pixels is not available. Fig. 5(a) and (d) show holes of forward interpolated frames from the 33th frame of *silent* and 3th frame of *mobile*. The irregular black regions illustrate holes.

In order to handle holes, the proposed algorithm uses the forward and backward interpolated frames to reduce the number of holes and therefore, improves the frame quality. Fig. 5(b) and (e) illustrate the effect of holes following combination of forward and backward interpolated frames. After the combination of forward and backward interpolated frames, the interpolated frame may still contain unknown pixels that need to be filled. We replace remaining unknown pixels with results obtained from the bilateral interpolated frame to fill the holes, as shown in Fig. 5(c) and (f). To handle overlaps, we use simple averaging from the overlapped pixels.

The proposed algorithm make a jointing of forward and backward interpolated frames, $f_i$, as follows,

$$f_i(i,j) = \begin{cases} f_f(i,j) & \text{if } f_b(i,j) = H \text{ and } f_f(i,j) \neq H \\ f_b(i,j) & \text{if } f_b(i,j) \neq H \text{ and } f_f(i,j) = H \\ \frac{1}{2}(f_f(i,j) + f_b(i,j)) & \text{if } f_b(i,j) \neq H \text{ and } f_f(i,j) \neq H \\ f_{bi}(i,j) & \text{otherwise} \end{cases} \quad (6)$$

where $H$ is a constant that indicate the pixels are within holes and $(i,j)$ represents the pixel location. If one pixel has only one existing value, because there is a hole in either forward interpolated frame, $f_f$, or backward interpolated frame, $f_b$, this existing value is taken as the ultimate value in final interpolated frame, $f_i$. If one pixel has two existing values from $f_f$ and $f_b$, the proposed algorithm applies the average of the two existing values as the ultimate value in the final interpolated frame. If there is no existing value from the $f_f$ and $f_b$, the proposed algorithm applies existing pixel in the same position in the bilateral interpolated frame, $f_{bi}$, as the ultimate value in the final interpolated frame, as shown in Fig. 5(c) and (f).

Finally, the proposed algorithm make the final interpolated frame, $f_u$, defined as,

if $BlockSAD \geq AdaptiveThreshold$

$$f_u(i,j) = \frac{1}{2}(f_{bi}(i,j) + f_i(i,j))$$

else

$$f_u(i,j) = \frac{1}{3}(2 \times f_{bi}(i,j) + f_i(i,j)) \quad (7)$$

where *BlockSAD*, represents the SAD of two candidate blocks of the previous and the next frames which their averages are replaced with the desired interpolated block, and the *adaptive threshold* is defined as the average of the *BlockSAD*s of blocks which are located before the desired interpolated block.

If *BlockSAD* of desired interpolated block is considered to be large, the desired interpolated block is assumed unreliable and quality of final interpolated frame is degraded. We use *adaptive threshold* as a comparison criterion, to improve the results. If the *BlockSAD* is larger than *adaptive threshold*, we set the desired interpolated block to average of the bilateral, forward, and backward interpolated frames. Otherwise, we set the desired interpolated block to weighted average of these frames. In this weighted average, we assign more weight to bilateral interpolated frame.

## III. EXPERIMENTAL RESULTS

In order to evaluate the performance of the proposed algorithm, we conducted experiments using twelve standard Common Intermediate Format (CIF) test sequences. All of the test sequences have a frame rate of 30 frames per second. We considered the first 102 frames of each test sequence. To obtain quality of the interpolated frames, we removed 50 odd frames (from frame 3 till frame 101) and interpolated them using 51 even frames. Then the peak signal-to-noise ratio (PSNR) values were computed from the difference between the original odd frames and the interpolated frames. Considerably, Stefan sequence just included 90 frames and so its PSNR is based on 44 frames. For experiments, in unilateral motion estimation, we set the block size to $8 \times 8$ pixels and the search range to $\pm 16$ pixels. Temporal symmetry is one of the characteristics of bilateral motion estimation, so large search range always leads to artifacts in the interpolated frame where some blocks in the scene are replaced with the background. Hence, conventional algorithms which use bilateral motion estimation, need to limit the search range with a smaller range. For instance, in [21], [24], the block size is $16 \times 16$ and the search range is $\pm 8$. Therefore, we set the block size to $16 \times 16$ pixels and the search range to $\pm 8$ pixels for bilateral motion estimation. For OBMC, each block is enlarged from $16 \times 16$ to $20 \times 20$ to reduce block artifacts.

### A. Subjective evaluation

We compared interpolated results of the proposed algorithm with bilateral method and unilateral method (jointing of forward and backward methods). Fig. 6 and Fig. 7 illustrate interpolated frames of the test sequences, *Stefan* and *Tennis*. We created weighted combination of unilateral and bilateral methods, based on adaptive threshold which yields a better results than each of unilateral and bilateral method alone. If bilateral method had better results in some blocks, its weight is higher than the weight of unilateral method.

### B. Objective evaluation

Table I illustrates the average PSNRs of 50 frames that interpolated by unilateral method, bilateral method and proposed algorithm. As seen in the Table, the proposed algorithm using combination of unilateral and bilateral methods yield a 1dB and 0.75dB average PSNR improvement compared with unilateral and bilateral method, respectively.

In order to evaluate the performance of the proposed algorithm, we compared it to the conventional unilateral FRUC [17] and conventional bilateral FRUC [21] which the test conditions were the same as in our experiments.

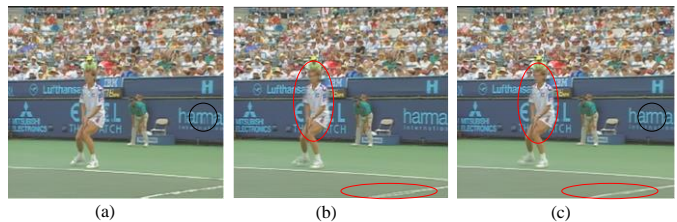

(a) (b) (c)

Fig. 6. Interpolated frames of frame 61 *Stefan* using (a) Unilateral method (PSNR = 28.07). (b) Bilateral method (PSNR = 28.27). (c) Proposed method (PSNR = 28.92). As showed in the figure, black circles demonstrate improvement the quality of the frame in (c) in comparison to (a), and red circles demonstrate the same improvement in (c) in comparison to (b).

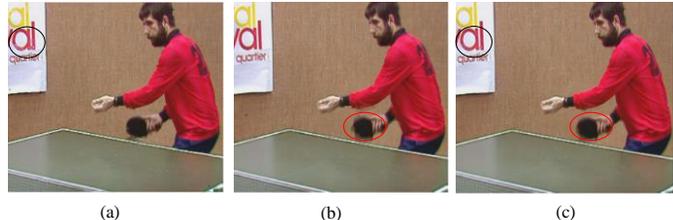

(a) (b) (c)

Fig. 7. Interpolated frames of frame 59 *Tennis* using (a) Unilateral method (PSNR = 27.92). (b) Bilateral method (PSNR = 28.45). (c) Proposed method (PSNR = 28.94). As showed in the figure, black circles demonstrate improvement the quality of the frame in (c) in comparison to (a), and red circles demonstrate the same improvement in (c) in comparison to (b).

TABLE I
AVERAGE PSNR OBTAINED USING THE UNILATERAL METHOD, BILATERAL METHOD AND PROPOSED ALGORITHM

| Test sequences | unilateral | bilateral | Proposed algorithm |
|---|---|---|---|
| *Foreman* | 33.30 | 33.47 | **34.17** |
| *Football* | 21.92 | 22.00 | **22.39** |
| *Mobile* | 24.94 | **28.82** | 26.79 |
| *Flower* | 29.52 | 29.79 | **30.40** |
| *Stefan* | 27.14 | 27.46 | **28.03** |
| *Coastguard* | 30.02 | **32.16** | 32.02 |
| *Paris* | 32.32 | 32.16 | **33.13** |
| *Soccer* | 28.71 | 29.14 | **29.72** |
| *Tennis* | 28.78 | 28.67 | **29.32** |
| *Akiyo* | 44.00 | **45.14** | 45.02 |
| *News* | 35.36 | 36.15 | **36.38** |
| *Silent* | 36.00 | 36.04 | **36.64** |
| *Average* | 31.00 | 31.75 | 32.00 |

Table II presents the average PSNRs of the three algorithms. As seen in Table II that our algorithm achieves an improvement of more than 0.7dB and 1.5dB than the conventional unilateral and bilateral FRUC, respectively.

Fig. 8 indicates the PSNRs for the interpolated frames in four test sequences: *Foreman*, *Flower*, *Silent* and *Stefan*. The graphs indicate the proposed algorithm are effective for entire frames.

TABLE II
AVERAGE PSNR OBTAINED USING THE CONVENTIONAL UNILATERAL FRUC, CONVENTIONAL BILATERAL FRUC AND PROPOSED ALGORITHM

| Test sequences | Conventional unilateral FRUC [17] | Conventional bilateral FRUC [21] | Proposed algorithm |
|---|---|---|---|
| *Foreman* | 32.71 | 31.82 | **34.17** |
| *Football* | 22.17 | 21.31 | **22.39** |
| *Mobile* | 26.18 | 25.39 | **26.79** |
| *Flower* | 27.93 | - | **30.40** |
| *Coastguard* | 30.75 | - | **32.02** |
| *Akiyo* | **45.11** | - | 45.02 |
| *News* | **37.25** | - | 36.38 |
| *Silent* | 35.81 | - | **36.64** |
| *Average* | 32.24 | - | 32.98 |
| *Average* | - | 26.17 | 27.78 |

## IV. CONCLUSION

In this paper, we proposed a novel MC-FRUC algorithm. The proposed algorithm combined unilateral and bilateral methods based on adaptive threshold to build a high quality of interpolated frames for MC-FRUC. Since motion-compensated frame interpolation along unilateral motion trajectories yields holes, this paper presented a new algorithm to fill holes. Experimental results showed that the proposed algorithm performed superior than existing algorithms both in terms of subjective and objective criterion evaluations.

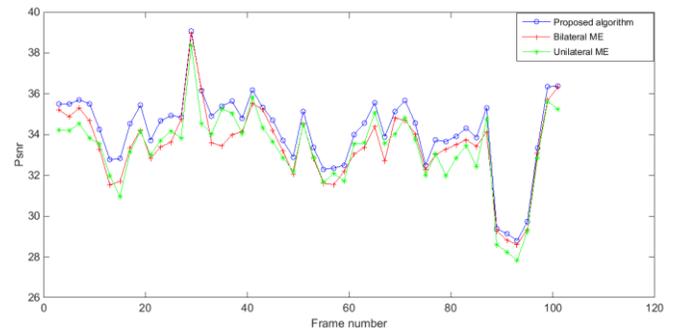
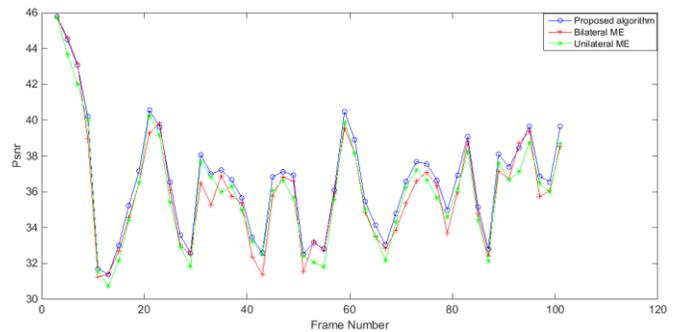
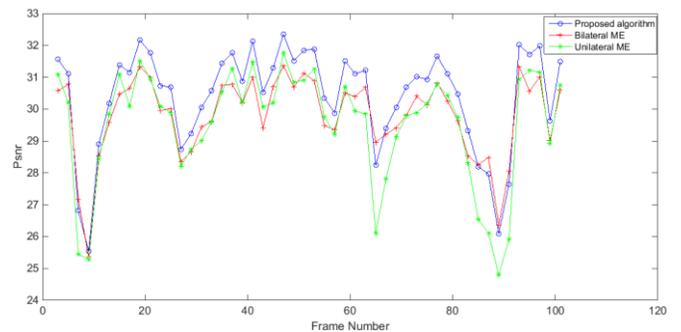
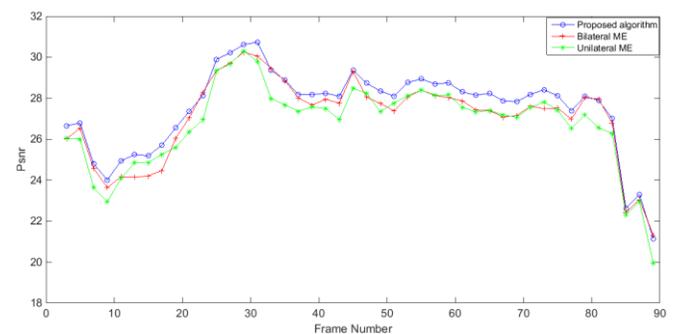

Fig. 8. PSNRs by the unilateral method, the bilateral method and the proposed algorithm for test sequences: (a) *Foreman*, (b) *Flower*, (c) *Silent* and (d) *Stefan*.